\documentclass[twocolumn,showpacs,preprintnumbers,amsmath,amssymb]{revtex4}

\usepackage{graphicx}
\usepackage{dcolumn}
\usepackage{bm}
\usepackage{float}
\usepackage{natbib}
\usepackage{amssymb}
\usepackage{amsmath}
\usepackage{epsfig}

\begin{document}
\title{Nonclassical photon pairs from warm atomic vapor using a single driving laser}

\author{L. Podhora, P. Ob\v{s}il, I. Straka, M. Je\v{z}ek, and L. Slodi\v{c}ka$^{*}$}

\address{Department of Optics, Palack\'y University, 17. listopadu 1192/12, 771~46 Olomouc, Czech Republic}

\email{slodicka@optics.upol.cz}


\begin{abstract}
	Generation of nonclassical light is an essential tool for quantum
	optics research and applications in quantum information
	technology. We present realization of the source of nonclassically
	correlated photon pairs based on the process of spontaneous
	four-wave-mixing in warm atomic vapor. Atoms are excited only by a
	single laser beam in retro-reflected configuration and narrowband
	frequency filtering is employed for selection of correlated photon
	pairs. Nonclassicality of generated light fields is proved by
	analysis of their statistical properties. Measured parameters of
	the presented source promise further applicability for efficient
	interaction with atomic ensembles.
\end{abstract}

\maketitle

\section{Introduction}

The scalable quantum-communication protocols based on atom-light
interfaces~\cite{DLCZ,Sangouard2011} have stimulated the
development of a number of experimental platforms aiming for
production of single photons with high purity, efficiency, and
frequency linewidth comparable to atomic natural
linewidth~\cite{White2016, Farrera2016, Fulconis2005,
	Somaschi2016, Higginbottom2016, Kuhn2002, Albrecht2013,Chu2016}.
Laser cooled
atoms~\cite{MOT_Du,Harris_Mot_single_freq,MOT_Kimble,Farrera2016}
have proven to be viable systems for achieving this goal either by
employing conditional parametric process or by exploiting the
intrinsic purity of single trapped atoms or ions as single photon
sources. The high level of isolation from thermal environment
achievable in these systems strongly enhances the ability of
photon generation with high purity and narrow linewidths. At the
same time, it is the necessity of isolation what makes these
systems spatially bulky and technically demanding. Recent
demonstrations of nonclassical photon pairs generation circumvent
these issues by utilization of warm atomic ensembles and several
excitation lasers in specially designed optical
pumping~\cite{Shu2016} or ladder-type electronic level
schemes~\cite{Lee2016}.

We report on further simplification of these schemes by
experimental demonstration of generation of nonclassical photon
pairs in warm atomic ensembles by the process of spontaneous
four-wave-mixing (SFWM) using the excitation of atoms with single
laser frequency, as previously realized with cold
atoms~\cite{Harris_Mot_single_freq}. Our scheme is based on the
counter-propagating laser excitation of $^{87}$Rb vapor on D1-line
resonant with the 5S$_{1/2}(F=2)\rightarrow$5P$_{1/2}(F=2)$
transition. Scattered Stokes and anti-Stokes photons are detected
at small angles with respect to the excitation beam, and the
polarization and frequency of generated photons are filtered using
pairs of Glan-Thompson polarizers and Fabry-P\'{e}rot resonators,
respectively. The coupling into opposite single mode fibers
directly guarantees the proper selection of the phase-matched
modes from the atomic emission. The nonlinear interaction is
further enhanced by focusing the excitation laser beam.
Uncorrelated photons coming from the initial population of the
\textit{F} = 2 ground state are suppressed to large extent by the
optical pumping mechanism happening for atoms before they enter
the observation area, which can be set to be slightly smaller than
the area covered by excitation laser beam~\cite{Shu2016}.


SFWM in atomic ensembles employs coherent collective spin
excitation of a large number of atoms. In the case of warm atomic
vapors, the coherence is typically strongly deteriorated by the
atomic motion. Coherent excitation takes place only on timescales,
where the recoil energy from the first Raman process cannot move
the atom out of the area given by its thermal de Broglie
wavelength~\cite{Mitchell_scat_dynamics}. For the case of Rb atoms
at 300\,K and right-angle scattering geometry, this would limit
the time for the generation of anti-Stokes photon to
about~130\,ps, which is extremely fast compared to typical
achievable Rabi frequencies in free space experiments, and even
for 2~degrees angle between excitation and observation directions
it gives the upper limit on the re-excitation time of 5\,ns. In
addition, fast thermal motion causes drift of atoms with
population in other than desired ground state into the interaction
area. In the scheme employing double-$\Lambda-$energy atomic level
structure, this leads to the emission of uncorrelated photons.
Recently, this has been circumvented by employment of atomic
polarization preserving coatings of vapor cells in combination
with additional optical pumping beam~\cite{Shu2016,Zhu2017} or
employment of other than double-lambda atomic energy level
schemes~\cite{Guo_ladder_config,Lee2016,Willis2010}. In addition
to these dominant effects, the large thermal motion of emitting
particles can cause their escape from the observation area and
contribute to decoherence of the spin wave due to change of mutual
atomic positions.

\section{Experimental implementation}

\begin{figure*}[!ht]
\centerline{\includegraphics[width=15cm]{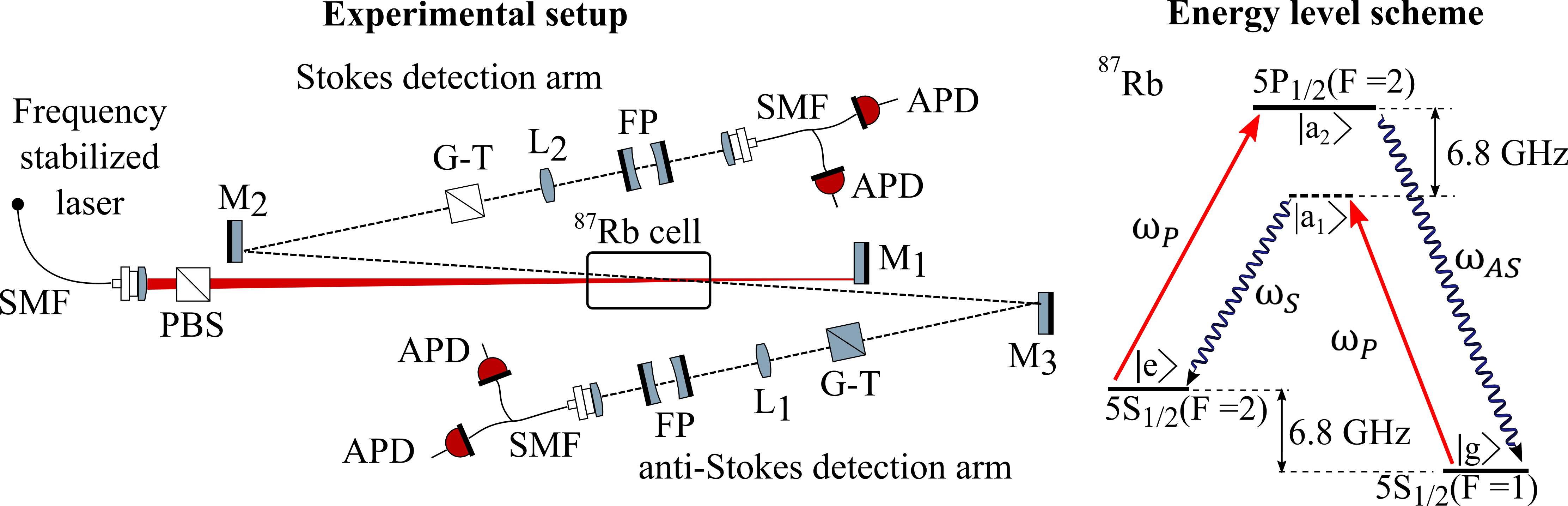}}
	\caption{The employed experimental setup and
		energy level scheme. Laser beam with frequency
		locked to resonance with the $|e\rangle \leftrightarrow|a_{2}\rangle$ transition passes polarization beam splitter (PBS) and
		excites $^{87}$Rb atoms in retro-reflected configuration. Its focus position is set precisely to coincide with the mirror plane (M$_1$). The photons
		emitted under small angle from the spatial region close to the
		cell back-face are polarization and frequency filtered using pairs
		of Glan-Thompson polarizers (G-T) and Fabry-P\'erot etalons (FP). The
		lenses (L$_1$, L$_2$) select and collimate the generated
		photons to optimize their coupling into opposite single-mode fibers (SMF)
		which guide them to the detection setups consisting of
		single-photon detectors (APDs).}
	\label{fig:block_scheme}
\end{figure*}

The limitations caused by high
temperature of atomic cloud can be partially overcome by realization of the SFWM process on short
timescales using high optical power density, frequencies close to
respective transition resonances, and employment of small-angle-scattering geometry~\cite{Mitchell_scat_dynamics}. In the presented experiment, the 7.5~cm long cylindrical glass cell filled with $^{87}$Rb is used as the nonlinear interaction medium. The cell does not contain any atomic polarization preserving coatings or buffer gas. The optical windows of the cell are antireflection coated to reduce spurious light
scattering when working under small observation angles. The large
ground state hyperfine splitting of $^{87}$Rb isotope gives
the advantage of easier spectral separation of generated photons. At
the same time, employing the D$_1$ spectral line with only two
excited states reduces off-resonant scattering from other than desired levels
and simplifies the theoretical description of the SFWM process.

The requirement of using a single laser frequency suggests the double-$\Lambda$
scheme as the only viable choice of energy level scheme. The respective transition strengths on D1-line lead to the excitation scheme depicted in the Fig.~\ref{fig:block_scheme}). We resonantly excite the $|e\rangle \sim 5S_{1/2}(F=2) \leftrightarrow
|a_{2} \rangle \sim 5P_{1/2}(F=2)$ transition which
results in the SFWM process starting by Raman transition from
$|g \rangle\sim 5S_{1/2}(F = 1)$ over $|a_{1}\rangle$ virtual level to $|e\rangle$ upon emission of the Stokes photon.
This is followed by fast re-excitation to $|a_{2} \rangle$ and
emission of anti-Stokes photon resonant with $|a_{2} \rangle
\leftrightarrow |g \rangle$ transition. 
The virtual energy level is given by the ground hyperfine energy level splitting and
ac-Stark effect from the excitation laser and therefore it is about
6.8\,GHz red-detuned from the $|a_{2}\rangle$ state.
Thermal atoms entering the excitation beam are optically
pumped into the ground state $|g \rangle$. This should ideally
happen before they enter the SFWM observation region, defined by
the overlap between the excitation and detection spatial modes, so
the uncorrelated background coming from atoms which could not
undergo Stokes emission is minimized~\cite{Shu2016}. 

A significant advantage of our particular experimental approach is
the great simplification in experimentally achieving the optimal FWM phase
matching spatial arrangement. In the case of employing the double-$\Lambda$
energy level scheme where frequencies of interacting photons are
nearly degenerate, the correlated photons are radiated close to
exactly opposite directions. The use of the single excitation
laser in retro-reflected configuration then allows for easy
alignment of the whole setup without any need for auxiliary seed
beams as in schemes employing several different wavelengths of
absorbed and emitted photons~\cite{Willis2010,guo}. The frequency difference of $\Delta\nu_{S,AS}=13.6$\,GHz between Stokes and anti-Stokes photons together with the cancellation of the excitation laser momenta in the back-reflection excitation configuration leads to the residual phase mismatch of the realized SFWM process.
This phase mismatch imposes the lower limit on the emission angle between excitation and observation directions for which the biphoton generation will be still efficient, given by the spatial length $d=c/(2 \Delta\nu_{S,AS})\sim$~11\,mm corresponding to the $\pi$ phase mismatch. Therefore, the observation angle should be large enough so that the excitation and observation spatial modes overlap only on the distance shorter than $d$. In the presented experiment, the observation angle of 2.3$^\circ$ together with observation and emission mode diameters of about 0.3~mm correspond to the length of the effective mode overlap of $2\times3.8$\,mm. We consider here the effective mode overlap defined as the overlap of two transverse Gaussian mode functions higher than 50\,\%.
We note, that alternative energy level scheme on $^{87}$Rb D1 line with similar
efficiency can be employed by excitation of $|g
\rangle\leftrightarrow|a_{2}\rangle$ transition with the virtual
energy level in the first Raman process being then blue-detuned by
about 6.8\,GHz from the $|a_{2}\rangle$ state, however
experimentally observed SFWM efficiency in this regime has been
similar or slightly lower.

\begin{figure}[!ht]
	\centerline{\includegraphics[width=8cm]{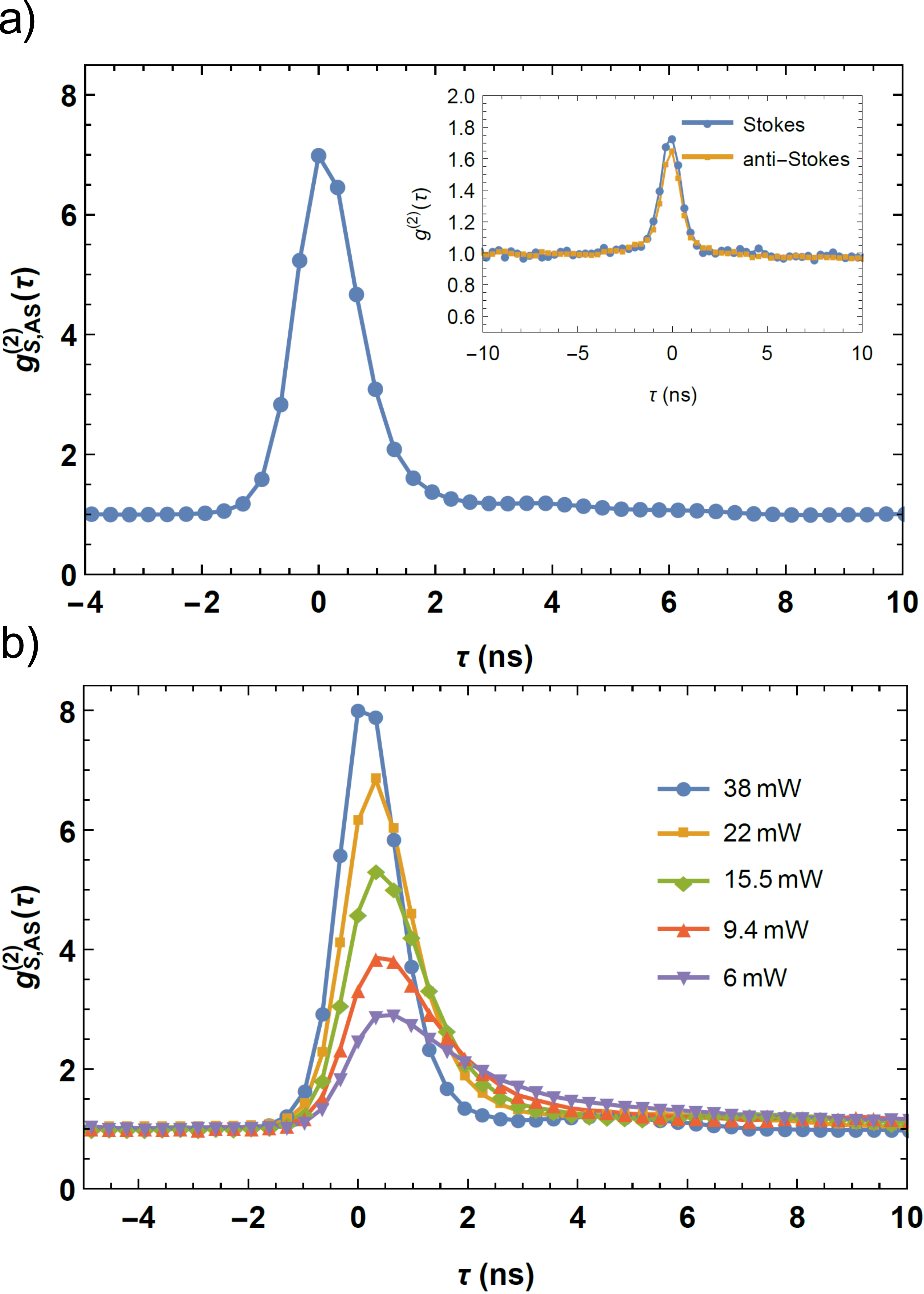}}
	\vspace{-0.5cm}
	\caption{a) Second-order correlation function between Stokes and anti-Stokes optical fields $g^{(2)}_{S,AS} (\tau)$. The inset shows the second-order correlations on individual Stokes  $g^{(2)}_{S,S}(\tau)$ (blue circles) and anti-Stokes $g^{(2)}_{AS,AS}(\tau)$ (yellow squares) modes. The error bars are too small to be visible on the scales of the presented plots. b) Measured power dependence of the second-order correlations $g^{(2)}_{S,AS} (\tau)$. }
	\label{fig:CS}
\end{figure}

The excitation laser beam coming from single mode fiber (SMF) has
polarization set to horizontal, that is parallel with respect to
the plane defined by optical table. The beam is focused so that
its waist coincides with the mirror (M1) behind the cell with the
full width at half maximum (FWHM) at the position of photon
generation of 0.3~mm. This narrows the interaction area and
increases the effective pump power density. The beam waist
positioned at the mirror ensures that the spurious effects of beam
focusing on the fulfillment of spatial phase matching conditions
are minimized. In the employed small-angle scattering geometry
with the angle of 2.3$^\circ$ between excitation and observation
light directions, the Stokes and anti-Stokes fields need to be
separated from the uncorrelated light scattered at strong $|e
\rangle \rightarrow |a_{2} \rangle $ transition. We use a pair of
Fabry-P\'{e}rot (FP) etalons centered at respective Stokes $|a_{1}
\rangle \rightarrow |e\rangle$ and anti-Stokes $|a_{2} \rangle
\rightarrow |g \rangle$ transitions. The particular setting of the
FP filters thus also sets the time order of the correlations
between the detected field modes. The etalons have free spectral
range of about 30~GHz, transmission linewidth (FWMH) of $\sim
0.6$\,GHz and peak transmissivity at 795~nm of 68\,\%. Additional
frequency filtering takes place in the $^{87}$Rb cell itself. As
can be seen in Fig.~\ref{fig:block_scheme}, the spatial alignment
is set such that the point of intersection between the excitation
and observation modes is close to the right end-face of the atomic
cell. This allows resonant anti-Stokes photons to leave the atomic
ensemble without much losses in the direction of anti-Stokes
detection setup. At the same time, far off-resonant Stokes photons
are passing the whole cell without attenuation while photons
resonant with $|e\rangle \rightarrow |a_{2}\rangle$ and
$|a_2\rangle \rightarrow |g\rangle$ transitions are strongly
attenuated in the same direction. In addition to uncorrelated
scattering processes, the excitation beam passes a number of
optical interfaces which are partially reflective and cause
reflections in the directions of observed photons. We install a
pair of Glan-Thompson (GT) polarizers set to transmit linear
polarization orthogonal with respect to excitation beam in each
observation direction to suppress the detection of these
reflections. The spatial mode of emitted photons is defined by the
combination of lenses (L$_1$ and L$_2$), fiber couplers, and
single mode fibers (SMF), which set the spatial width of the
observation region to 290~$\mu$m. The overall coupling efficiency
of the auxiliary aligning laser from Stokes to anti-Stokes fiber
is 11.5~\%, limited mostly by the finite transmission of FP
cavities. Single-mode fibers in both channels are connected to
detection setups consisting of single avalanche photodiodes (APDs)
or Hanbury-Brown-Twiss arrangements.

\section{Results}

The nonclassical properties of generated light fields are
investigated by measurement and
evaluation of normalized second-order correlation function defined as
$g^{(2)}_{n,m} (t_1,t_2)  = \langle a_n^\dagger (t_1) a_m^\dagger(t_2)
a_m(t_2) a_n(t_1) \rangle /(\langle a_n^\dagger a_n\rangle
\langle a_m^\dagger a_m\rangle)$,
where $a^\dagger$ and  $a$ are the creation and annihilation operators, $n,m$ are field indexes and $t_1, t_2$ are photon detection times. The nonclassicality of generated two mode
correlations can be revealed by comparison of Stokes - anti-Stokes
photon correlations with the correlations measured on individual
modes. For all classical states of the optical field, their mutual
relation is bound by the Cauchy-Schwarz inequality (C-S)
$[g^{(2)}_{S,AS} (t_{1}, t_{2})]^{2} \leq  g^{(2)}_{S,S}
(t_{1},t_{1}) g^{(2)}_{AS,AS} (t_{2},t_{2})$,
where subscripts $S$ and $AS$ denote measurement of second-order correlation
between Stokes and anti-Stokes channels, respectively. We quantify the violation
of the C-S inequality by defining the violation factor $F = [g^{(2)}_{S,AS}
(\tau)]^{2}/(g^{(2)}_{S,S} (0) g^{(2)}_{AS,AS}(0))$, where $\tau=t_2-t_1$
is time between photon detections in $S$ and $AS$ channels.

\begin{figure}[!ht]
	\centerline{\includegraphics[width=8cm]{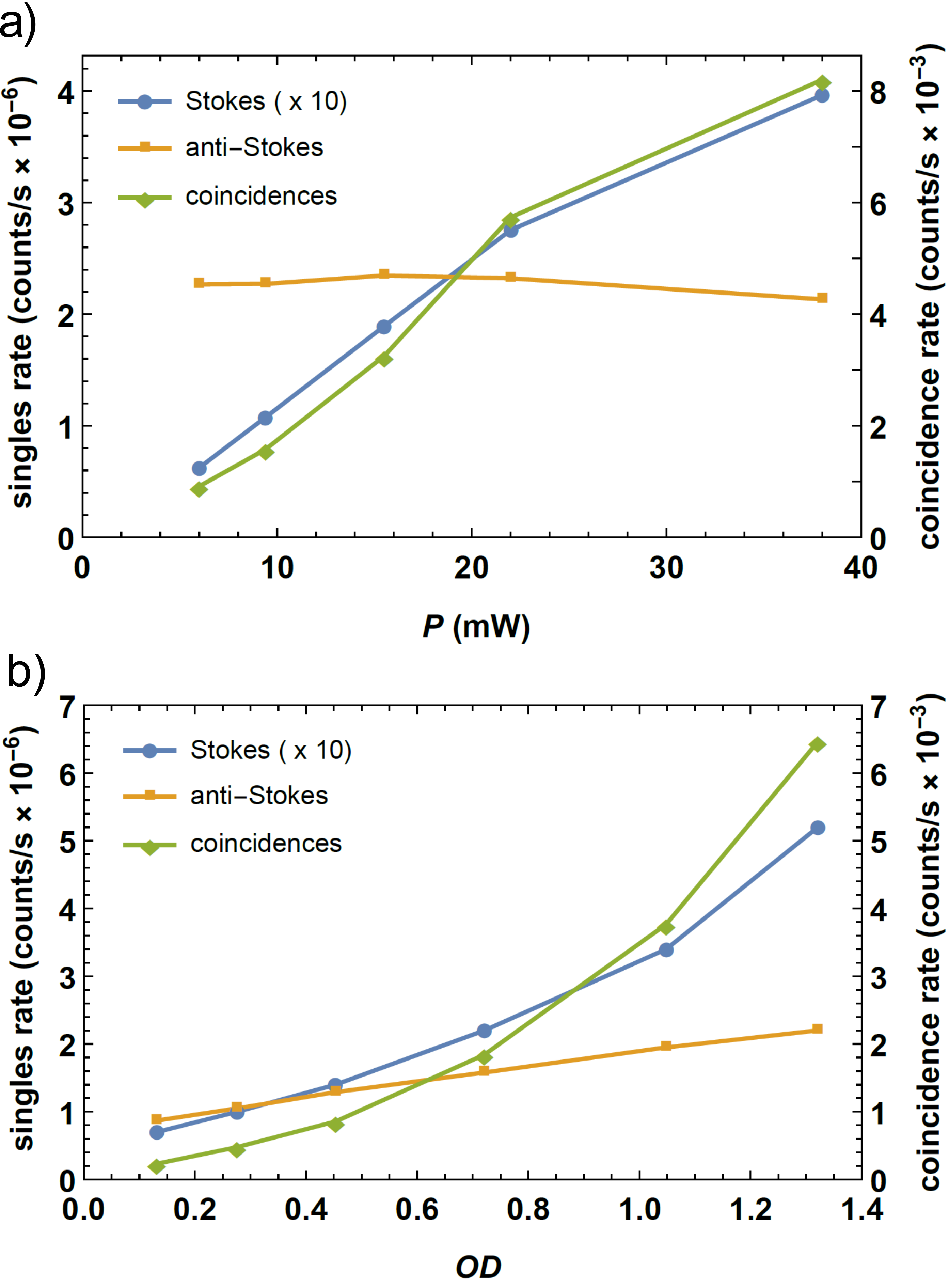}}
	\caption{The measured rates of Stokes and anti-Stokes singles and coincidences detections as a function of power a) and optical density b).} \label{fig2}
\end{figure}

Typical results of correlation measurement for optical depth of~1.3 and excitation beam power of 40~mW corresponding to 75\,minutes of acquisition time are shown in the Fig.~\ref{fig:CS}-a).
The maximum of $g^{(2)}_{S,AS}(0)=(6.984 \pm 0.004)$ and the maxima of autocorrelation functions $g^{(2)}_{S,S}(0)=(1.73 \pm 0.02)$ and $g^{(2)}_{AS,AS}(0)=(1.649 \pm 0.009)$ correspond to the violation of C-S inequality by factor $F = (17.1\pm 0.2)$. The excitation laser frequency detuning from the $|e\rangle\rightarrow |a_{2}\rangle$ transition has been optimized for maximization of the observed two-photon correlation $g^{(2)}_{S,AS}(0)$. The resulting optimal value corresponds to 70~MHz blue-detuned from the unperturbed $|e\rangle\rightarrow |a_{2}\rangle$ transition, which is much smaller than the Doppler-broadened spectral width of the atomic transition of  $\sigma_{\rm D} \sim 230$\,MHz. This is mainly due to the requirement on the fast read-out of the stored excitation after the process of Stokes photon scattering, while the multi-photon Stokes excitations are naturally suppressed due to strongly off-resonant excitation of the Stokes transition in the presented single-laser scheme.
Measurement of $g^{(2)}_{S,AS}(\tau)$ shows almost symmetric
correlation peak caused by high excitation rate on the resonant
$|e\rangle\rightarrow |a_{2}\rangle$ transition together with
thermal motion of atomic cloud, which leads to broadening of the
frequency bandwidth of emitted fields to scales larger than
linewidths of employed Fabry-P\'erot resonators. The resonators
thus filter the detected time envelope to broader and symmetrical
profile. The FWHM temporal width of the correlation
$g^{(2)}_{S,AS}(\tau)$ has been evaluated to $1.3$~ns, what is in
good agreement with the theoretically expected value calculated by
convolution of spectral widths of the two FP filters and specified
time jitter of employed APDs of 350~ps. The peak values of
autocorrelation measurements in Stokes and anti-Stokes modes
suggest a partially thermal statistics of the detected light
fields. The statistics of these individual modes should be
generally indistinguishable from the one of the thermal field
\cite{Blauensteiner_thermal_statistics,Yurke_Potasek}, however,
for the observed temporally short correlation functions, the
finite time jitter of employed avalanche photodiodes decreases the
correlation values substantially. Imperfect single mode nature of
generated photons and small fraction of photons coming from other
than desired SFWM parametric process also likely contribute to
suppression of observed photon bunching. The FWHM temporal widths
of correlations are equal to 1.3~ns and 1.2~ns for Stokes and
anti-Stokes fields, respectively. All presented correlation
functions were evaluated for 0.32~ns coincidence windows.

The detected biphoton rate within the FWHM temporal width of the S-AS correlation
peak of $1.3$~ns  is $N_{S,AS} = (7224\pm 85)$~pairs/s what together with
Stokes and anti-Stokes count rates of $N_{S}=4.8\times
10^5$~counts/s and $N_{AS}=18.9\times 10^5$~counts/s corresponds to
overall triggering efficiency of anti-Stokes photons of
$\eta=N_{S,AS}/ N_{S} = 1.51\,\%$. The rate of biphotons per 1~mW of the excitation laser power is about 180~pairs/(s mW) and the estimated biphoton rate for the triggered anti-Stokes photon bandwidth of 1\,MHz is about 9.4 pairs/sec. The high ratio of anti-Stokes to Stokes count rates is caused by the detection of photons
resonant with $|a_{2}\rangle\rightarrow|g\rangle$ transition,
which do not come from the SFWM process, but are simply scattered
by atoms being already in the excited state $|e\rangle$. This
points to imperfect optical pumping caused by the similarity of
the spatial widths of excitation and observation regions. The
excitation beam diameter was not much larger than the observation
area due to the limited available laser power in our experiment
and, consequently, the trade-off between achievable Rabi frequency
at the beam center and the efficiency of the optical pumping
mechanism.

The crucial parameter influencing the efficiency and the time scale of
SFWM interaction is optical power density of the exciting laser
beam. Fig.~\ref{fig:CS}-b) shows the second-order
correlations measured for different excitation beam powers.
The rest of the experimental settings is kept the same as in the correlation measurement shown in Fig.~\ref{fig:CS}-a). The correlation maxima of $g^{(2)}_{S,AS}$ depend approximately linearly on the laser power, however, this dependence slowly tends to saturate for high excitation powers due to the increase of uncorrelated noise photons~\cite{Shu2016}. The measured power dependence shows, that lower excitation powers give longer excitation times on the $|e\rangle\rightarrow|a_{2}\rangle$ transition manifested in delayed start of the whole process. At the same time, this leaves more space for decoherence of the atomic spin wave, which becomes observable already at few nanoseconds time scales. The lower excitation power also results in longer biphoton time
envelope and ability to resolve the expected exponential decay on anti-Stokes transition, as the emitted biphoton bandwidth becomes smaller than linewidths of employed FP filters. Due to the power limitations of our laser source, we were not able to further explore the scaling and seemingly emerging saturation of the power dependence. Each measured correlation curve here corresponds to 5\,min acquisition time. We note, that the slightly higher observed maximum of $g^{(2)}_{S,AS} (\tau)$ correlation at 38\,mW excitation power compared to measurement presented in Fig.~\ref{fig:CS}-a) is caused by the different spatial coupling efficiency of the biphotons into the single mode detection channels after realignment of the experimental setup.

We further characterize the observed SFWM process by measurement of the dependence of generated photon rates on the excitation laser power and atomic density while keeping the rest of the experimental settings the same as in the correlation measurement presented in Fig.~\ref{fig:CS}-a). The figure~\ref{fig2}-a) shows the expected linear dependence of far-detuned Stokes photon rates on the excitation laser power. Together with the constant emission rate of saturated anti-Stokes transition, this results in linear increase of the measured S-AS coincidences. The figure~\ref{fig2}-b) shows the dependence of detected photon rates on optical depth set by the temperature of atomic cell. An observed quadratic increase of coincidence rate suggests that the anti-Stokes generation process is based on collective superradiant emission. Each data point in Fig.~\ref{fig2}-a) and b) corresponds to 300\,s and 60\,s acquisition time, respectively.

\begin{figure}[!t!]
	\centerline{\includegraphics[width=8cm]{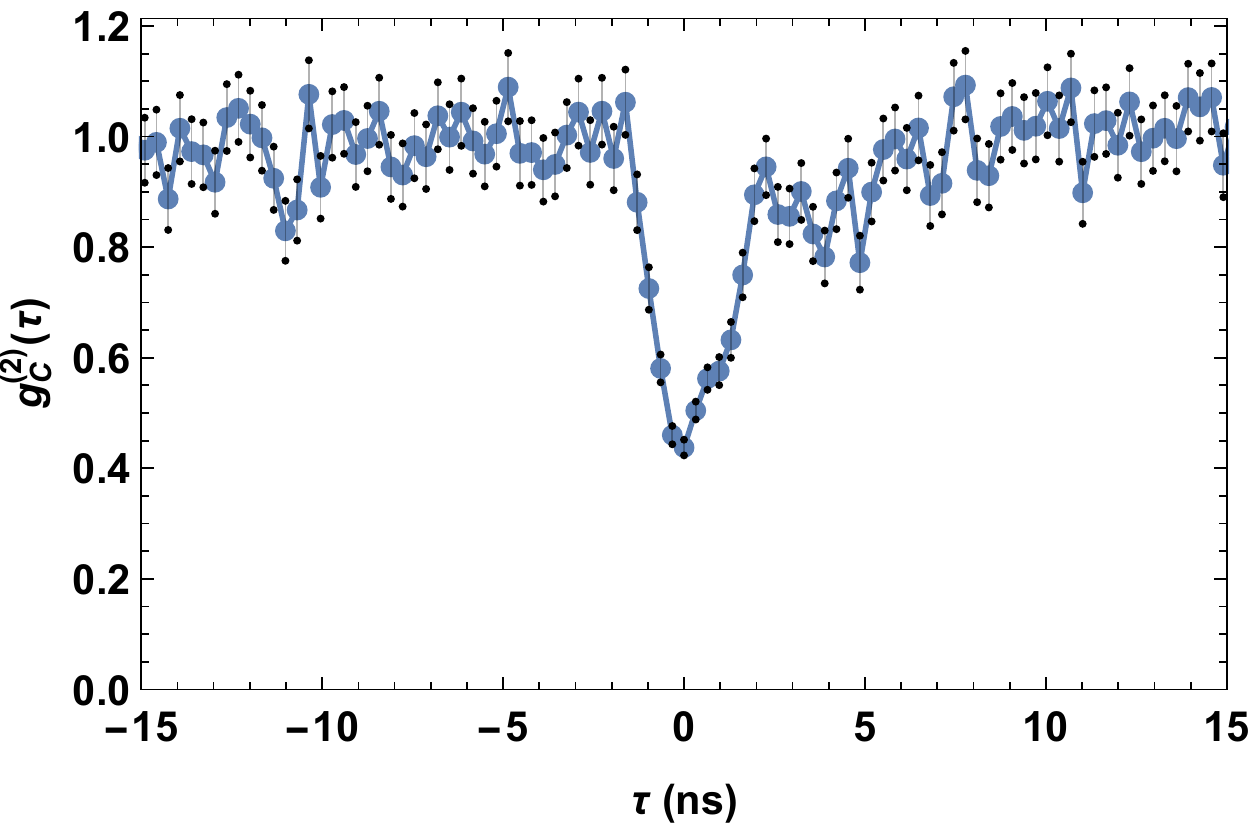}}
	\caption{Measured conditional correlation function $g^{(2)}_{C}(\tau)$. The error bars correspond to a single standard deviation.}
	\label{fig:Triggered_g2_plot}
\end{figure}

The strong violation of the Cauchy-Schwarz inequality proves the
nonclassical correlations between the two light fields, however,
it does not reveal any information about the state of individual
light modes conditioned on the detection of photon in the other
mode. We implement the measurement of the normalized conditional
second-order correlation function on the
anti-Stokes field conditioned on the detection of a Stokes
photon~\cite{Grangier,Lee2016}, which can be evaluated as $g^{(2)}_{C}(\tau)=N_{S,AS1,AS2}(\tau) N_S/(N_{S,AS1}(0) N_{S,AS2}(\tau))$, where $N_{S,AS1,AS2}(\tau)$ is the histogram of three-photon coincidences composed of the two-fold coincidence counting event between the Stokes and anti-Stokes (AS1) detectors with fixed delay which maximizes the $g^{(2)}_{S,AS1}$ correlation and the single counting event on the second anti-Stokes detector (AS2) for various delays $\tau$. The $N_S$ corresponds to the total number of triggering Stokes detections, $N_{S,AS1}$ is the total number of two-photon coincidences for the time delay corresponding again to the $g^{(2)}_{S,AS1}$ correlation maximum and $N_{S,AS2}(\tau))$ is the histogram of two-photon coincidences between the Stokes and anti-Stokes (AS2) detectors. The measured function presented in
Fig.~\ref{fig:Triggered_g2_plot} shows a clear anti-bunching
property of the triggered field with the minimal value of
$g^{(2)}_{C}(0)=0.44\pm0.01$. This further proves the heralded
presence of nonclassical light and in addition, it suggests, that
the mean photon number of the heralded state is $\langle n \rangle
< 2$. The employed experimental settings are the same as in the
measurements of S-AS correlations presented in
Fig.~\ref{fig:CS}-a) and the total acquisition time was
75~minutes. The FWHM temporal width of the triggered anti-Stokes
photons was evaluated to 2.7~ns, caused mainly by the measurement
of three-fold coincidences which correspond to time convolution of
three photon wave-packets. The minimum value of $g^{(2)}_{C}(0) =
0.44\pm 0.01$ corresponds very well to the measured correlations
$g^{(2)}_{S,AS} (\tau)=6.984\pm 0.004$ if we consider that the
detection post-selected state in the Stokes and anti-Stokes arms
can be well approximated by $|1_S,1_{AS}\rangle$ state. With this
assumption, the ratio of photons coming from the desired SFWM
process to uncorrelated noise photons corresponds to theoretical
value of $g^{(2)}_{C}(0) = 2 g^{(2)}_{AS,AS}(0)/g^{(2)}_{S,AS}(0)
= 0.47$,  where factor two corresponds to the degeneracy of
detection probabilities at time delays shorter than time envelopes
of detected AS photons.

\section{Conclusion and outlook}

We have presented the experimental realization of nonclassical
light source based on the process of SFWM in warm atomic vapor and
employing of only single excitation laser. The nonclassical
properties of generated light fields have been proven by violation
of Cauchy-Schwarz inequality by factor $17.1 \pm 0.2$ and by
measurement of the anti-Stokes field statistics conditioned on the
Stokes photon detection resulting in $g^{(2)} (0) = (0.44 \pm
0.01)$. The measured length of tunable temporal correlations
on the order of several nanoseconds together with the biphoton generation rate of more than $7.2\times10^3$ pairs/s suggest a direct applicability of the presented source for interaction with target atomic ensembles~\cite{Mercadier2009,Kaczmarek2017,Wolters2017}, but also leave space for the further frequency filtering with the prospect of linewidths on the order of few MHz and with biphoton rates still applicable for investigation of storage in narrow-band atomic quantum memories~\cite{Eisaman2005, Novikova2012}.

Compared to the recent realizations of nonclassical sources with warm
atomic vapors~\cite{Guo_ladder_config,guo,Shu2016,Zhu2017,Willis2010},
the overall experimental scheme has been strongly simplified.
Using the single excitation laser not only reduces the complexity
and spatial demands of the whole experiment, but also allows for simple alignment of the spatial
phase-matching in of SFWM process. Further enhancement of the
purity of generated single photons and their generation rate is
expected by increasing power of the excitation laser beam, which
has been limited in the current setup to 40\,mW. This would allow
a broader spatial width of the excitation beam at the interaction
region, what would lead to better optical pumping while
maintaining the required high speed of the anti-Stokes photon
generation. The scheme allows for independent optimization of the
Stokes and anti-Stokes transitions excitation probabilities by
attenuation of the back-reflected excitation laser, what will also
further improve observed nonclassical
correlations~\cite{guo,Zhu2017}. The presented results are likely
to contribute to the development of quantum repeater based
communication architectures with spatially small and technically
simple single-photon sources. Furthermore, the demonstrated scheme
has a strong potential for application in many experimental platforms
as a technically simple source of heralded photons in close analogy to the extensively used source of broad-band photons
based on spontaneous parametric down-conversion.

\section*{Acknowledgements}
This work was supported by the Grant Agency of Czech Republic GB14-36681G and Palack\'y University IGA-PrF-2017-008.
We thank Morgan Mitchell, Robert Sewell, Ferran Martin, Micha\l{} Parniak and Gabriel H\'etet for valuable discussions.



\end{document}